\documentclass[twocolumn,prc,floatfix,showpacs,preprintnumbers,%
superscriptaddress,nofootinbib]{revtex4}
\usepackage[utf8x]{inputenc}
\usepackage{mathrsfs}
\usepackage{amssymb}
\usepackage{amsmath}
\usepackage{graphicx}

\usepackage{xcolor}
\definecolor{lcolor}{rgb}{0.5,0,0}
\definecolor{citcolor}{rgb}{0,0.3,0.0}

\usepackage[breaklinks,colorlinks,urlcolor=blue,citecolor=citcolor,linkcolor=lcolor]{hyperref}
\usepackage{mciteplus}

\newcommand{\der}{\mathrm{d}}
\newcommand{\rt}{{\mathbf{r}_T}}
\newcommand{\rtsqr}{{\mathbf r}^2_{T}}

\newcommand{\bt}{{\mathbf{b}_T}}
\newcommand{\btsqr}{{\mathbf b}^2_{T}}

\newcommand{\nc}{{N_\mathrm{c}}}

\newcommand{\gev}{\ \textrm{GeV}}

\newcommand{\as}{\alpha_{\mathrm{s}}}

\newcommand{\eq}{Eq.~}

\newcommand{\sigmaa}{{ \sigma^A_\textrm{dip} }}

\newcommand{\sigmadip}{{ \sigma_\textrm{dip} }}

\newcommand{\xpom}{{x_\mathbb{P}}}

\newcommand{\A}{{\mathcal{A}}}

\begin{document}

\author{Heikki Mäntysaari}
\affiliation{
Department of Physics, %
 P.O. Box 35, 40014 University of Jyväskylä, Finland
}
\affiliation{
Physics Department, Brookhaven National Laboratory, Upton, NY 11973, USA
}

\author{Raju Venugopalan}
\affiliation{
Physics Department, Brookhaven National Laboratory, Upton, NY 11973, USA
}

\title{
Systematics of strong nuclear amplification of gluon saturation from exclusive vector meson production in high energy electron-nucleus collisions
}


\preprint{}

\begin{abstract}
We show that gluon saturation gives rise to a strong modification of the scaling in both the  nuclear mass number $A$ and the virtuality $Q^2$ of the  vector meson production cross-section in exclusive deep-inelastic scattering off nuclei. We present qualitative analytic expressions for how the scaling exponents are modified as well as quantitative predictions that can be tested at an Electron-Ion Collider. 

\end{abstract}

\maketitle

\section{Introduction}

A striking observation made by the HERA deep inelastic scattering (DIS) experiments on electron-proton collisions is the rapid growth of the gluon densities at high energies or, equivalently, at small values of longitudinal momentum fraction $x$ carried by the gluons~\cite{Aaron:2009aa,Abramowicz:2015mha}. Weak coupling QCD studies predict that when gluon occupancies become of order $1/\as$, where $\as$ is the QCD coupling constant, the perturbative bremsstrahlung of gluons is balanced by strong repulsive gluon self-interactions. This dynamics generates a phenomenon called gluon saturation that tempers the growth in gluon distributions~\cite{Gribov:1984tu,Mueller:1985wy}. For every $Q^2\gg \Lambda_{\rm QCD}$, with $\Lambda_{\rm QCD}$ representing the intrinsic QCD scale, there is an $x$ value for which the maximal occupancy is reached. Equivalently, at small $x$, for each value of $x$ there exists saturation scale $Q_{s,p}^2(x)$ in the proton; transverse momentum modes below this scale have maximal occupancy. High momentum modes $k_\perp \gg Q_{s,p} (x)$ asymptote to the usual perturbative QCD dynamics. The underlying physics of the onset of and the many-body dynamics of gluon saturation is captured in a Color Glass Condensate (CGC) effective theory~\cite{McLerran:1993ni,McLerran:1993ka,McLerran:1994vd,Gelis:2010nm,Kovchegov:2012mbw,Blaizot:2016qgz} which allows for efficient computations.  

Despite the large amount of precise HERA data, the evidence for saturation effects is unclear even though recent analyses increasingly point in this direction~\cite{Rezaeian:2012ji,Golec-Biernat:2017lfv}. One reason for this uncertainty is that the proton's saturation scale $Q_{s,p}^2 (x)$ is not very large at the $x$ values probed in the HERA experiments. Discussions of gluon saturation in hadron-hadron collisions can be found in \cite{JalilianMarian:2005jf,Lappi:2010ek}. The search for gluon saturation in particular and the many-body non-linear dynamics of gluons in general is a major motivation for the Electron Ion Collider proposal~\cite{Accardi:2012qut} in the US and the LHeC proposal at CERN~\cite{AbelleiraFernandez:2012cc}. 

The physics of gluon saturation is universal. All non-perturbative information specific to the quantum numbers of a given nucleus is encoded in the initial conditions; at asymptotically small $x$, at a fixed impact parameter, an external probe would be unable to distinguish between a heavy nuclear target and a proton. All memory of their initial conditions would be lost and their saturation scales would be nearly identical~\cite{Mueller:2003bz}. However for realistic values of $x$, as shown explicitly in the McLerran-Venugopalan (MV) model~\cite{McLerran:1993ni,McLerran:1993ka,McLerran:1994vd}, nuclear information is contained in the atomic number ($A$) -dependence of the saturation scale. While not universal in the larger sense of complete independence of the dynamics from the initial conditions with which the system is ``prepared", it is nonetheless remarkable that the dynamics of a nucleus at high energies is controlled primarily by a single $A$-dependent scale. In the MV model, simple kinematic and dynamical arguments suggest that $Q_{s,A}^2(x) \sim Q_{s,p}^2(x)\,A^{1/3}$; this expection is borne out in detailed models~\cite{Kowalski:2007rw} that we shall discuss further shortly. 

The search for gluon saturation will require that we identify measurements where its onset will show strikingly different systematics from those seen in its absence. 
Based on the arguments outlined, it is widely believed that effects of gluon saturation are likely precocious in high-energy ($e+A$) DIS off nuclei. We will demonstrate here that the onset of gluon saturation is especially dramatic in exclusive vector production off large nuclei. The simple reason why exclusive processes are attractive measurements to pursue is that the perturbative QCD cross-section in such processes is proportional to the nuclear gluon distribution \emph{squared}~\cite{Ryskin:1992ui}. Since the gluon distribution grows rapidly at small $x$, such measurements are especially sensitive when gluon distributions saturate.

We will discuss here the $Q^2$ and $A$ dependence of exclusive vector meson production in $e+A$ collisions within the CGC  framework. Exclusive production of vector mesons in this saturation picture of $e+A$ DIS has been studied previously in the  literature~\cite{Goncalves:2004bp,Kowalski:2006hc,Caldwell:2009ke,Lappi:2010dd,Toll:2012mb}. A recent development in exclusive vector meson electroproduction is the extraction of event-by-event fluctuations of gluon spatial distributions in the proton from HERA data~\cite{Mantysaari:2016ykx}.  These results are very relevant to understanding recently discovered ``ridge-like" multiparticle correlations in proton-proton and proton-nucleus collisions~\cite{Dusling:2015gta,Mantysaari:2017cni} and can also be studied in future in $e+A$ collisions~\cite{Kowalski:2008sa,Toll:2012mb}. There has been recent progress in developing the theory beyond leading log $x$ accuracy~\cite{Boussarie:2016bkq}; we expect that these developments will in future help us refine our estimates.

Our focus here will be to demonstrate that gluon saturation gives rise to very specific (and strong) dependencies in the $A$ and $Q^2$ scaling of exclusive vector meson cross-sections for $Q^2 > Q_{s,A}^2$ that are qualitatively different from those for $Q^2 < Q_{s,A}^2$, where $Q_s$ stands for the saturation scale of the target.  Exclusive processes off nuclei are currently being studied in in ultraperipheral heavy-ion collisions at RHIC and the LHC where one nuclei acts as a source of (quasi) real photons~\cite{Bertulani:2005ru,Klein:2017boo} for the other;  there are several recent works that explore the role of small-$x$ dynamics in these measurements~\cite{Goncalves:2011vf,Adeluyi:2012ph,Abelev:2012ba,Abbas:2013oua,Lappi:2013am,Khachatryan:2016qhq,Mantysaari:2017dwh}. However in these reactions, the photon virtuality is limited to be $Q^2\sim 1/R_A^2$, where $R_A$ is the nuclear radius. Further, studying the $A$ dependence by varying the heavy-ion beams is experimentally challenging. 

These limitations do not exist at the EIC~\cite{Accardi:2012qut} where exclusive processes can be measured over a wide range of $Q^2$, $A$ as well as rapidity separation $y_\mathbb{P}= \ln(1/\xpom)$ between the vector meson and the target. The variable $\xpom$ is the equivalent of Björken $x$ for an exclusive process; it has the parton model interpretation of being the momentum fraction of a parton within the colorless (``Pomeron") exchanged between the virtual photon probe and the nuclear target. The kinematical coverage of the EIC is discussed in detail in Ref.~\cite{Aschenauer:2017jsk}.

\section{Exclusive vector meson production}
At high energies, DIS can be formulated conveniently in a dipole picture, whereby the incoming photon with virtuality $Q^2$ splits into a quark-antiquark dipole and the dipole subsequently scatters elastically off the target with the amplitude $N$; in exclusive vector meson production, the dipole hadronizes into the vector meson.  The virtual photon to quark-antiquark dipole splitting amplitude $\Psi_{\gamma^* \to q\bar q}$ can be computed in quantum electrodynamics ~\cite{Kovchegov:2012mbw}. The dipole to vector meson hadronization amplitude $\Psi_{q\bar q \to VM}$ is non-perturbative. We therefore have to rely on phenomenological parametrizations of the vector meson wavefunction; we will employ here the Gauss-Light Cone parametrization from \cite{Kowalski:2006hc}. Because the time scales for hadronization are far greater than the time scale of the interaction of the dipole with the target, ratios of the exclusive vector meson production cross-sections in different nuclei should at most weakly depend on the the details of these non-perturbative wavefunctions.

The scattering amplitude for exclusive vector meson production can  be written as~\cite{Kowalski:2006hc}
\begin{multline}
\label{eq:diffamp}
\A = i\int \der^2 \rt \der^2 \bt \frac{\der z}{4\pi} [\Psi_{\gamma^*\to q\bar q}\Psi^*_{q\bar q \to VM}](\rt,z,Q^2) \\
\times e^{-i (\bt + (1-z) \rt) \cdot {\mathbf \Delta}} \frac{\der \sigmadip}{\der^2 \bt} (\bt,\rt,\xpom). 
\end{multline}
Here $z$ is the fraction of the momentum of the photon carried by the quark (the antiquark carries the remaining fraction $1-z$), 
${\mathbf \Delta}\approx \sqrt{-t}$ is the transverse momentum transfer from the target and $\sigmadip$ is the dipole-target cross-section. The difference between forward and non-forward wavefunctions gives the extra factor $\exp[ i(1-z)\rt \cdot {\mathbf \Delta}]$~\cite{Bartels:2003yj,Kowalski:2006hc}. We anticipate that this factor  has only a small effect at small $x$ and its role will be further diminished in the ratios we will examine and will not include it in our numerical calculations. The coherent diffractive vector meson production cross-section reads as 
\begin{equation}
\frac{\der \sigma^{\gamma^*A \to VA}}{\der t} = \frac{1}{16\pi} \left| \A \right|^2,
\end{equation}
where $V$ denotes the vector meson of interest.

In our numerical computations in the following sections, we will use the IPsat model to describe the dipole-proton cross-section $\sigmadip$. The IPsat model contains an eikonalized DGLAP-evolved gluon distribution function (see also Ref.~\cite{McLerran:1998nk}), which guarantees the correct perturbative limit in the dilute region (small $x$ and large $Q^2$) and preserves unitarity because $\der \sigmadip / \der^2 \bt \to 2$ for large dipoles. The impact parameter dependence is included by having the saturation scale depend on the transverse distance from the center of the proton. In the IPsat model, the expression for the dipole cross-section is~\cite{Kowalski:2003hm}
\begin{equation}
\frac{\der \sigmadip}{\der^2 \bt} = 2 \left[ 1 - \exp \left( -\frac{\pi^2}{2\nc} \rtsqr \as(\mu^2) xg(x, \mu^2) T_p(\bt) \right) \right],
\end{equation}
with the Gaussian proton transverse profile function:
\begin{equation}
T_p(\bt) = \frac{1}{2\pi B_p} e^{-\btsqr/(2B_p)}.
\label{eq:proton-profile}
\end{equation}
The free parameters in the model are the proton size  $B_p$ measured by the dipole probe and the initial conditions for the DGLAP evolution of the gluon distribution function $x\,g(x,Q^2)$. These were fixed in Ref.~\cite{Rezaeian:2012ji} by fitting the HERA inclusive DIS data. The dipole amplitude obtained was applied successfully previously in several phenomenological works~\cite{Tribedy:2010ab,Tribedy:2011aa,Lappi:2013am}. The dipole-proton cross-section can be generalized to nuclei as in Refs.~\cite{Kowalski:2003hm,Kowalski:2007rw,Lappi:2010dd}.  It is employed for instance to construct the IP-Glasma initial conditions for $A+A$ collisions~\cite{Schenke:2012wb,Schenke:2012fw,Gale:2012rq}. We also include the so-called skewedness and real part corrections as in Ref.~\cite{Lappi:2010dd}. These corrections are phenemenologically necessary in order to describe the HERA data\footnote{We emphasize that the skewedness correction becomes numerically large at large $Q^2$ (see Fig. 15 in \cite{Mantysaari:2016jaz});
 we do not consider the absolute cross-sections calculated in this work to be reliable in that kinematical domain.}. 
 These corrections do not affect the $A$ scaling in our discussion. They have  a small effect on the $Q^2$ scaling which mostly cancels in the cross-section ratios.

The generalization of the IPsat dipole cross-section to nuclear targets, following the procedure suggested in \cite{Kowalski:2003hm,Kowalski:2007rw}, is given by 
\begin{equation}
\label{eq:ipsat_nuke}
	\frac{\der \sigmaa}{\der^2 \bt} = 2\left[ 1- e^{ -\frac{A}{2} T_A(\bt) \sigmadip } \right],
\end{equation}
where $\sigmadip$ is the total dipole-proton cross-section integrated over the impact parameter.
The nuclear transverse density profile $T_A$ is obtained by integrating the Woods-Saxon distribution over the longitudinal coordinate, and is normalized as $\int \der^2 \bt T_A(\bt)=1$.

We will supplement our numerical study by simple analytical estimates that capture the underlying physics behind the $Q^2$ and $A$ scaling of the vector meson cross-sections. These analytical estimates are easier to realize in a simpler dipole model that is sufficient for this purpose, the Golec-Biernat--Wusthoff (GBW) model ~\cite{GolecBiernat:1998js}. In this model, the dipole-proton scattering amplitude can be written as 
\begin{equation}
\label{eq:gbw}
\frac{\der \sigmadip}{\der^2 \bt}  = 2\left[ 1-e^{-\rtsqr Q_{s,p}^2} \right] ,
\end{equation}
and the saturation scale $Q_{s,p}$ implicitly depends on $\xpom$. The dipole-nucleus cross-section is obtained from \eq~\eqref{eq:gbw} by replacing $Q_{s,p}^2 \to Q_{s,A}^2 \sim A^{1/3} Q_{s,p}^2$.

In the following, we will present numerical results computed in the IPsat model for $J/\Psi$ and $\rho$ production in the kinematics relevant for the EIC and LHeC as well as analytical estimates in asymptotic kinematics computed in the GBW model. An advantage of studying exclusive $\rho$ electroproduction is that, at moderate $Q^2$, the contribution from the saturation region is enhanced relative to the case of the heavier $J/\Psi$ meson. On the other hand, since the $\rho$ is much lighter than the $J/\Psi$, the lack of a large momentum scale makes perturbative computations questionable at smaller $Q^2$ values.

\section{Results for high  $Q^2: Q^2 > Q_{s,A}^2$.}
We will consider exclusive vector meson production in the high $Q^2$ region alone in this section and will take up the case of low $Q^2$ in the following section. In this high virtuality region, $Q^2$ is the only relevant scale and the mass difference between the $J/\Psi$ and the $\rho$ is less relevant. 
\subsection{$A$ scaling}
It is instructive to first consider the forward limit of $t=0$. Here, 
in the dilute (``pQCD") region of $Q_{s,A}^2 \ll Q^2$, we can expand all exponents in and \eqref{eq:gbw} and get
\begin{equation}
\A \sim \int \der^2 \rt \der^2 \bt [\Psi_{\gamma^*\to q\bar q}\Psi^*_{q\bar q \to VM}]\, \rtsqr Q_{s,A}^2
\end{equation}
where $Q_{s,A}^2\sim A^{1/3} Q_{s,p}^2$. 
Performing the $\bt$ integral, one gets another factor of $A^{2/3}$ from the area, which gives 
\begin{equation}
\A \sim A \int \der^2 \rt \Psi^*\Psi \rtsqr \,.
\end{equation}
Squaring the amplitude to obtain coherent cross-section, one obtains 
\begin{equation}
	\frac {\der \sigma^{\gamma^* p \to Vp}}{\der t}(t=0) \sim A^2 \,.
\end{equation}

\begin{figure}[tb]
			\includegraphics[width=0.5\textwidth]{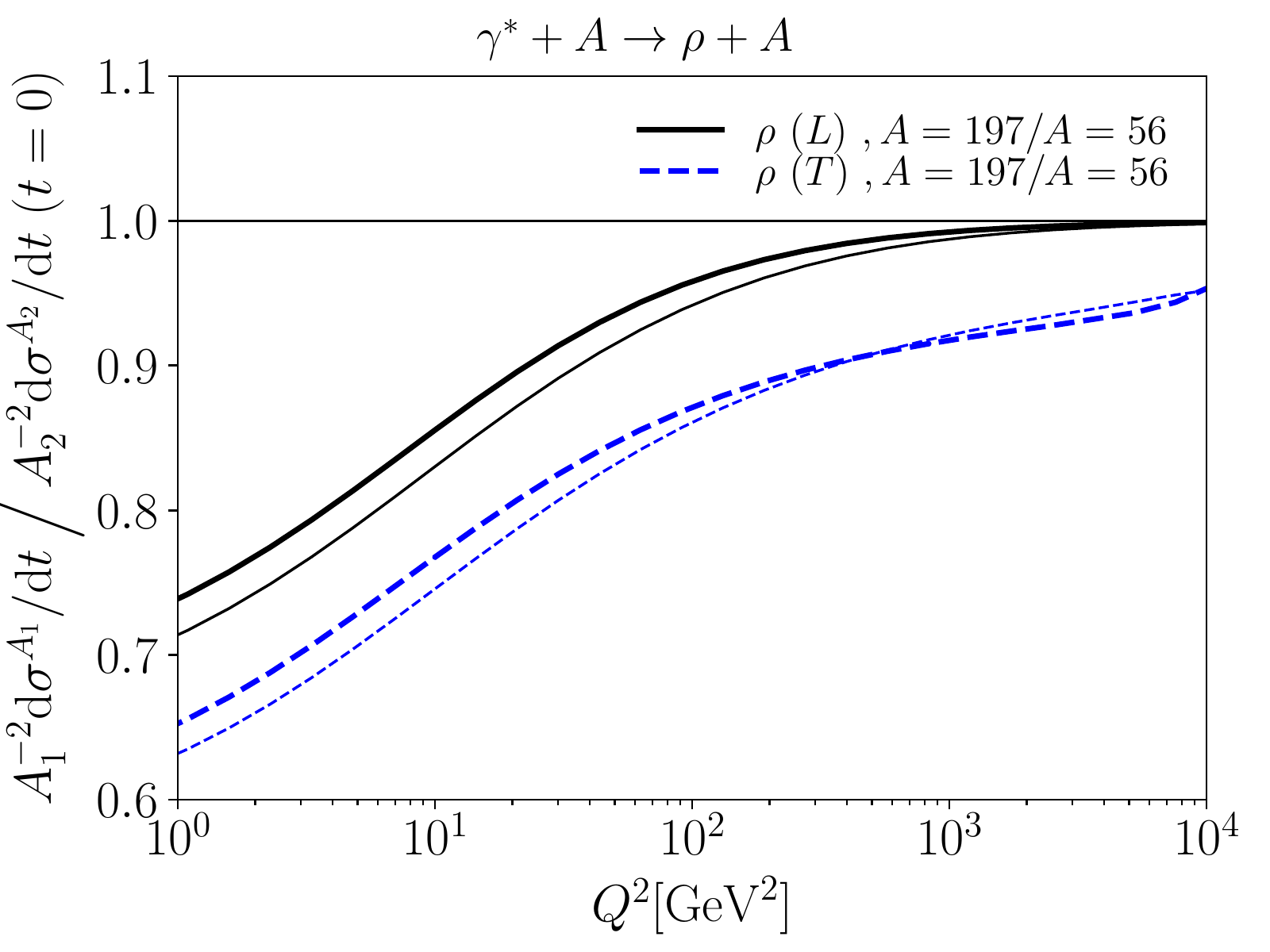} 
				\caption{The $A^2$ scaling at $t=0$ for $\rho$ production. $T$ and $L$ refer to transverse and longitudinal polarization, respectively. On the y-axis is plotted the $A^2$ normalized ratio of the cross-sections for Gold over Iron. Thick solid and dashed lines correspond to $\xpom=0.01$ and thin solid and dashed lines to $\xpom=0.001$.}
		\label{fig:Asqr_scaling_t0}
\end{figure}

This scaling for the IPsat model is demonstrated numerically in Fig.~\ref{fig:Asqr_scaling_t0} which shows the normalized ratio of the exclusive $\rho$ meson cross-sections at $t=0$ for Gold and Iron nuclei respectively. Such ratios are desirable because as noted previously model dependencies cancel; this is also the case for systematic uncertainties in the experiments.
Fig.~\ref{fig:Asqr_scaling_t0}  demonstrates that for longitudinal polarization, the asymptotic $A^2$ scaling is seen for $Q^2=10^3$ GeV$^2$, though it is within 5\% already for $Q^2=10^2$ GeV$^2$. At smaller $Q^2$, significant suppression seen is due to gluon saturation. The suppression is greater for transversely polarized photons which are more sensitive to larger dipole sizes. 
The energy or $\xpom$ dependence of the ratio is weak, with the suppression increasing only slightly when $\xpom$ is decreased from $0.01$ to $0.001$. 
For comparison, we note the values for the saturation scales\footnote{The quark saturation scale $Q_s^2$ is defined as $\frac{\der \sigmadip}{\der^2 \bt}\left(\rt^2=\frac{2}{Q_s^2}\right) = 2\left(1-e^{-1/2}\right)$.} of Gold and Iron nuclei at $b=0$ extracted from the IPsat model generalized to nuclei as shown in Eq.~\eqref{eq:ipsat_nuke}. For Gold, we get the value of the quark saturation scale to be $Q_{s,A}^2(x=0.01)=0.85\gev^2$ and $Q_{s,A}^2(x=0.001) = 1.2\gev^2$. The corresponding numbers for Iron are $0.54\gev^2 $ and $0.75\gev^2$ respectively.

\begin{figure}[tb]
\centering
		\includegraphics[width=0.5\textwidth]{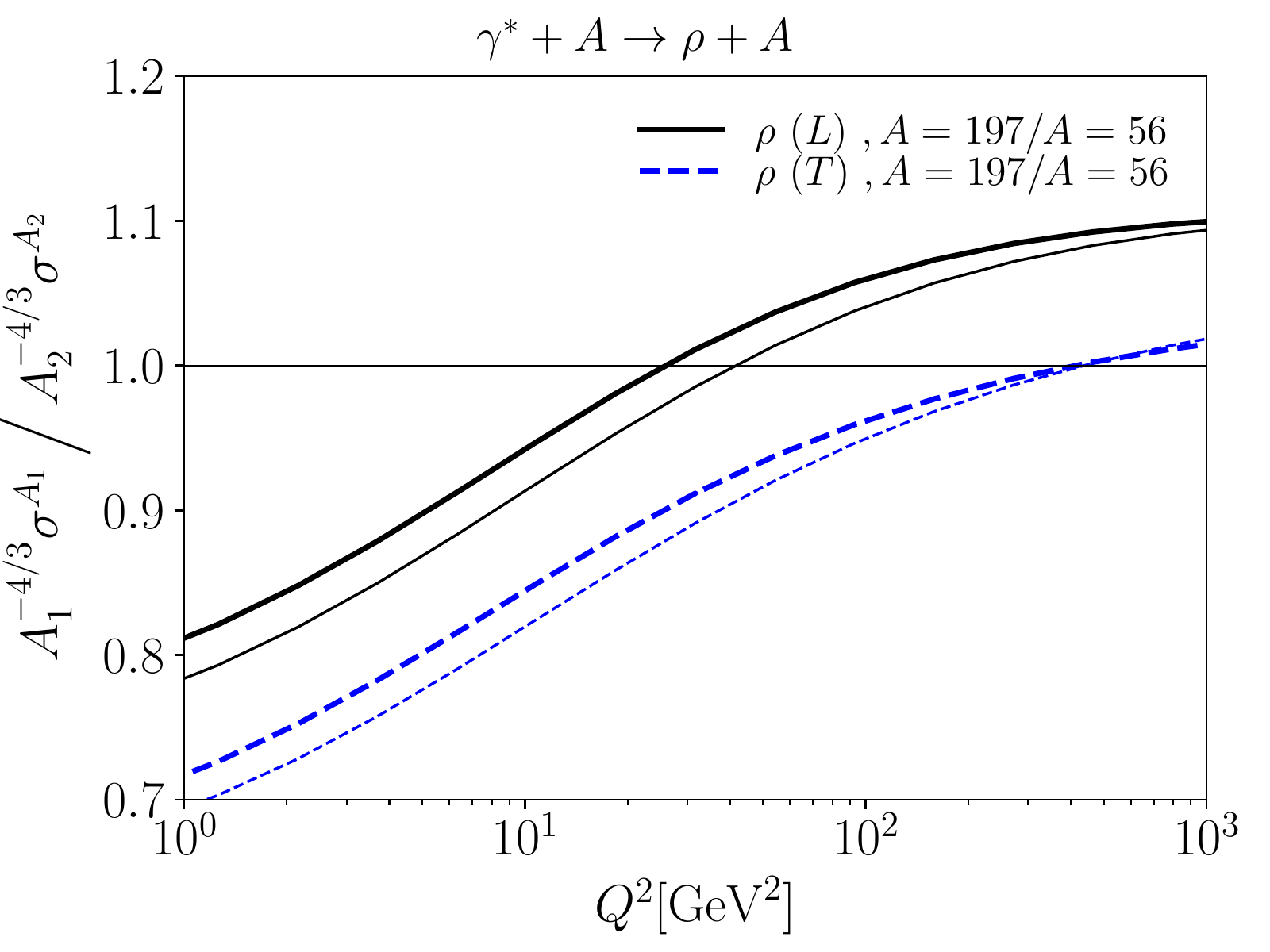} 
				\caption{The $A^{4/3}$ scaling for the total coherent $\rho$ production cross-section On the y-axis is plotted the $A^{-4/3}$ normalized ratio of the cross-sections for Gold over Iron. Thick lines correspond to $\xpom=0.001$ and thin lines $\xpom=0.01$.}
		\label{fig:A_scaling_totxs}
\end{figure}

Since the width of the coherent peak is $t_\text{max} \sim 1/R_A^2 \sim A^{-2/3}$, the total exclusive vector meson cross-section (integrated over $t$) scales like
\begin{equation}
	\sigma^{\gamma^*A \to VA} \sim A^2 A^{-2/3} = A^{4/3}\, .
\end{equation}
The numerical result for this scaling in the IPsat model is shown in Fig.~\ref{fig:A_scaling_totxs}. At high $Q^2$, the normalized ratio does not go to unity.  This is a consequence of oversimplifying the $t$ integral to include just the width of the coherent peak to give the factor $A^{-2/3}$, and is valid only at asymtotically large $A$. 

\begin{figure}[tb]
\centering
		\includegraphics[width=0.5\textwidth]{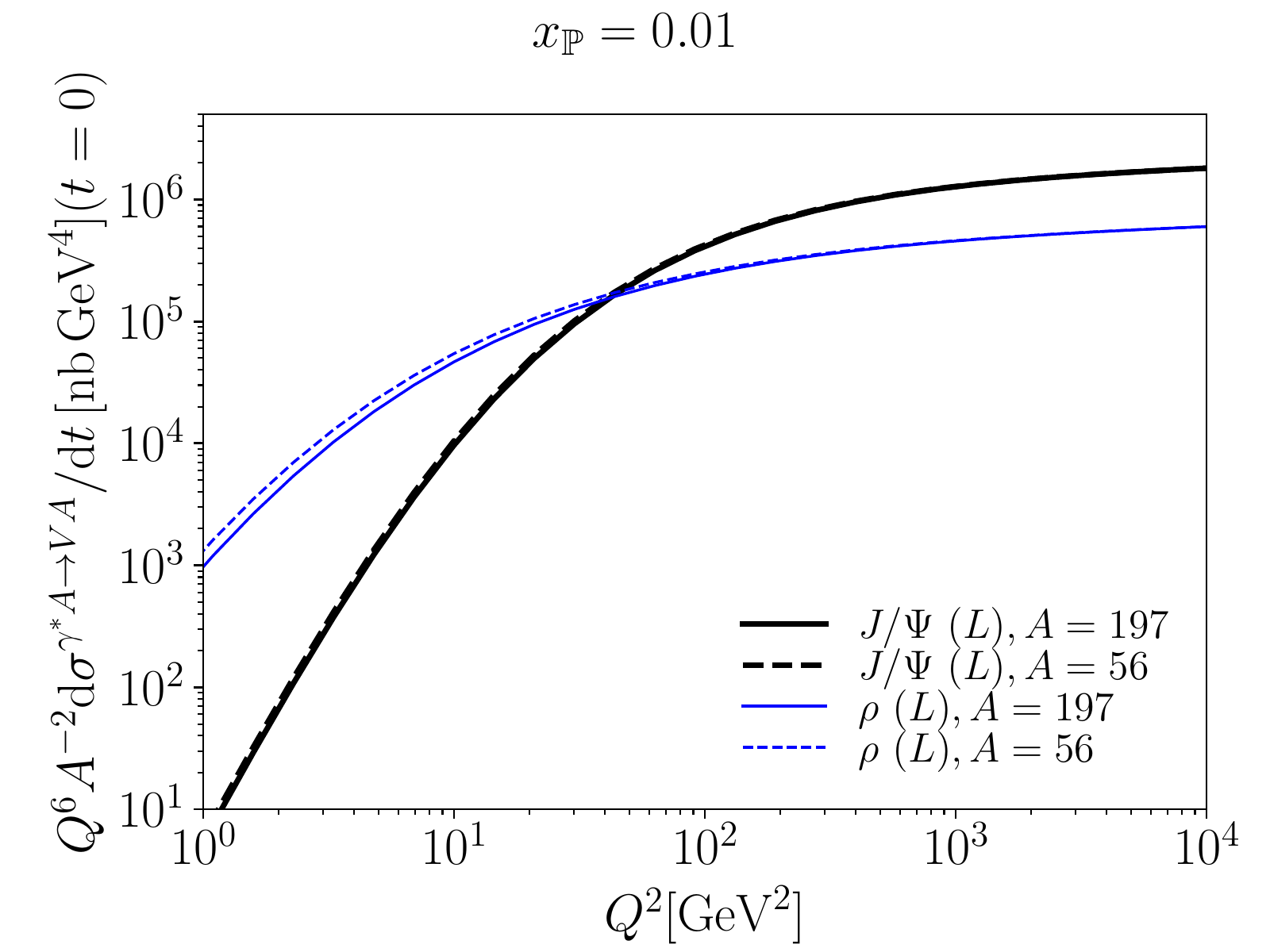} 
				\caption{The $Q^6$ scaling at $t=0$ for longitudinal vector meson production (thick black lines: $J/\Psi$, thin blue lines: $\rho$) in the dilute region. The exclusive cross-section meson cross-section multiplied by $Q^6$ flattens out at large $Q^2$.}
		\label{fig:Q6_scaling_t0}
\end{figure}

\subsection{$Q^2$ scaling}
For a given choice of the vector meson wavefunction (longitudinal Gauss-LC from Ref.~\cite{Kowalski:2006hc}), the overlap of this wavefunction with the longitudinally polarized photon wavefunction has the dependence 
\begin{equation}
	\left(\Psi_{\gamma^*\to q\bar q}\Psi^*_{q\bar q \to VM}\right)_L \sim z(1-z) Q K_0(\varepsilon r) \phi_L(r,z)\,,
\label{eq:long-wave-overlap}
\end{equation}
with $\varepsilon = \sqrt{Q^2z(1-z) + m_q^2} \approx Q$ for $Q^2 \gg Q_s^2$ and $|\rt|=r$. The scalar part of the vector meson wavefunction $\phi_L \sim z(1-z)e^{-r^2 M_V^2}$ limits contributions from dipoles larger than the inverse vector meson mass $M_V$.  A stronger limit on dipole sizes, $r\lesssim 1/Q$, is set by the Bessel function,  and the $Q^2$ scaling becomes
\begin{equation}
 \A_L \sim \int \der r \, r \,r^2 Q K_0(Qr) \sim Q \frac{1}{Q^4} \sim \frac{1}{Q^3}\, .
\end{equation}
Thus the exclusive longitudinal $\rho$ cross-section (both the total and its value at $t=0$) has the $Q^2$ scaling~\cite{Brodsky:1994kf}
\begin{equation}
	\sigma_L^{\gamma^*A \to VA} \sim |\A_L|^2 \sim \frac{1}{Q^6}\, .
\end{equation}
The numerical result for the dependence of the $\rho$ and $J/\Psi$ exclusive cross-section on $Q^2$ (scaled by $Q^6$) is shown in Fig.~\ref{fig:Q6_scaling_t0}. The $Q^{-6}$ behaviour of the cross-section at high $Q^2$ is apparent since the scaled cross-section is flat in the region $Q^2 \gtrsim 10^2 \gev^2$. The $A^2$ scaling is also visible, as the curves corresponding to different nuclei lie on top of each other in the $Q^2$ range plotted in the figure.

\begin{figure}[tb]
\centering
		\includegraphics[width=0.5\textwidth]{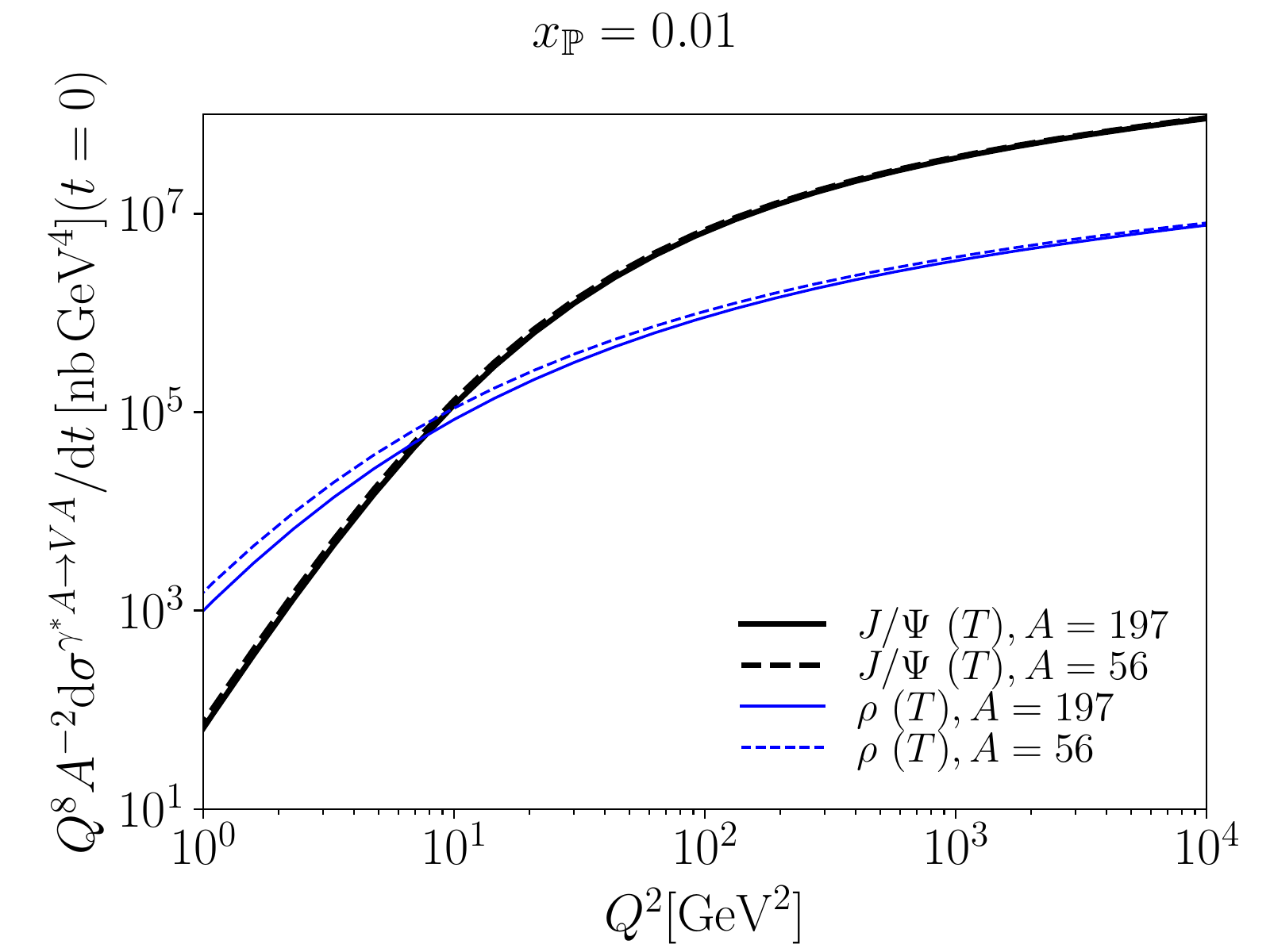} 
				\caption{The $A^2Q^{-8}$ scaling of the transverse vector meson production cross-section at $t=0$. }
		\label{fig:Q8_scaling_t0}
\end{figure}

For transversely polarized photons, we can perform a similar analysis as shown for the longitudinal polarization case in Eq.~\eqref{eq:long-wave-overlap}. Neglecting terms proportional to the light quark mass, the wavefunction overlap becomes
\begin{equation}
\left(\Psi_{\gamma^*\to q\bar q}\Psi^*_{q\bar q \to VM}\right)_T \sim \frac{1}{z(1-z)} \varepsilon K_1(\varepsilon r) \partial_r \phi_T(r,z)\,,
\end{equation}
where the scalar part of the vector meson wavefunction now behaves as $\phi_T(r,z)\sim z^2(1-z)^2e^{-r^2M_V^2}$. Thus the difference to the longitudinal polarization case corresponds to an extra power of $r$ in the scattering amplitude. The latter therefore scales as $Q^{-4}$ instead of $Q^{-3}$ and the  cross-section  (at $t=0$) correspondingly scales as
\begin{equation}
	\sigma_T^{\gamma^*A \to VA} \sim |\A_T|^2 \sim \frac{1}{Q^8}\, .
\end{equation}
We can check this analytical estimate numerically and the results are shown in Fig.~\ref{fig:Q8_scaling_t0}. We plot there the coherent $\rho$ and $J/\Psi$ production cross-section at $t=0$ scaled by $Q^8$. The proposed scaling from our simple argument  is not as accurate as in the case of longitudinal polarization. This is because the contribution for transversely polarized photons from large dipoles is not suppressed by large $Q^2$, a consequence of the strong dependence of the overlap wavefunction in this case on the $z\to 0,1$ limits. Even if the $Q^8$ scaling is not apparent in the $Q^2$ range accessible at the EIC, the large suppression of the exclusive vector-meson cross-section with $Q^2$ is striking and can be cleanly tested.

\section{Results for low $Q^2: Q^2 < Q_{s,A}^2$}

\begin{figure}[tb]
\centering
		\includegraphics[width=0.5\textwidth]{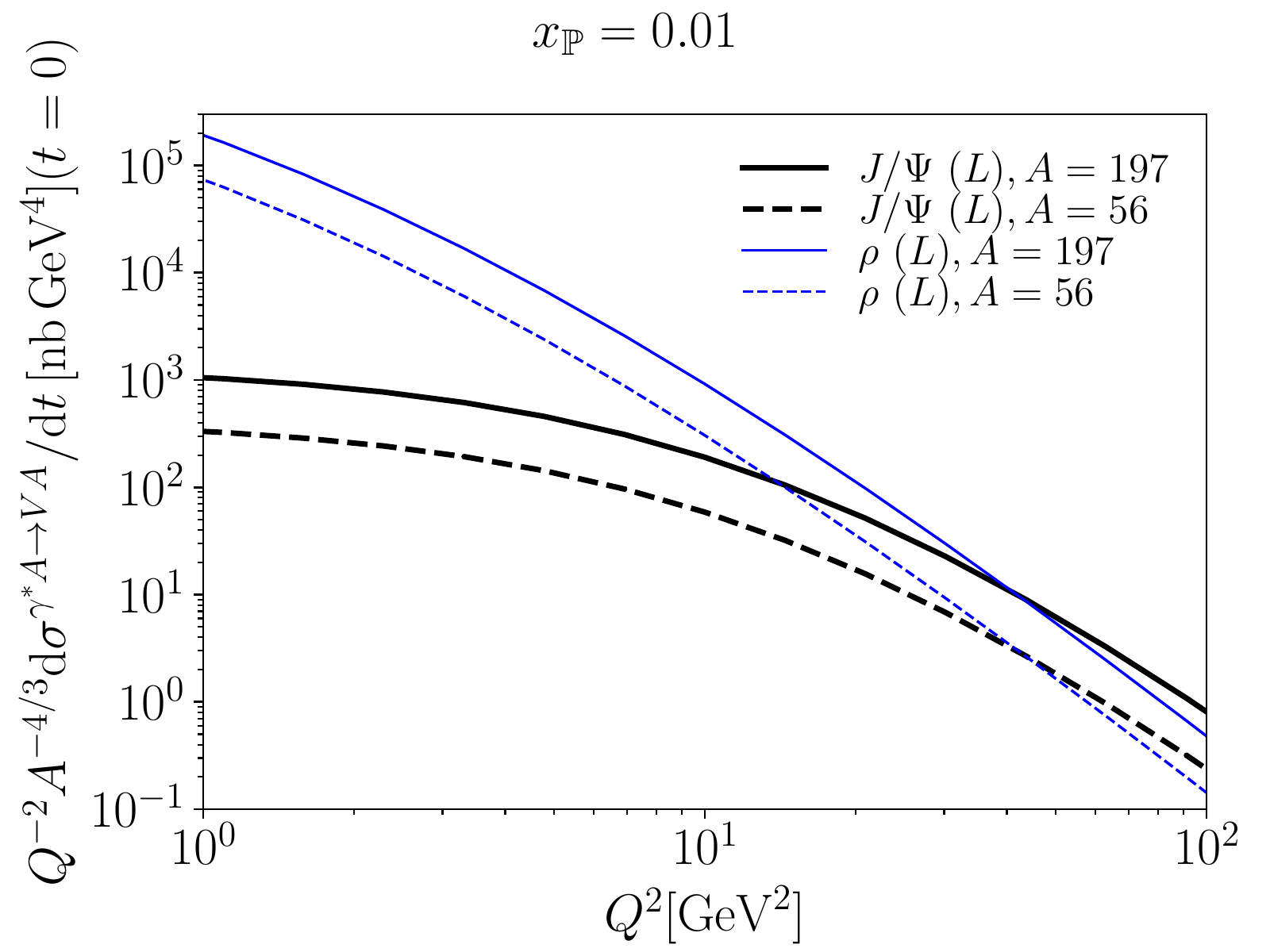} 
				\caption{The cross-section for coherent longitudinal vector meson production at $t=0$. At low $Q$, the cross-section is flat at low $Q^2$ when scaled by $Q^{-2}$.  For $\rho$, this behavior is only obtained at asymptotically small $Q^2$ values where our model is not applicable.The  cross-section is scaled in $A$ by the analytical asymptotical expectation $\approx A^{4/3}$. Our result here shows that the  scaling is not exact for realistic kinematics.}
		\label{fig:lowQ_scaling_Qdep}
\end{figure}

\begin{figure}[tb]
\centering
		\includegraphics[width=0.5\textwidth]{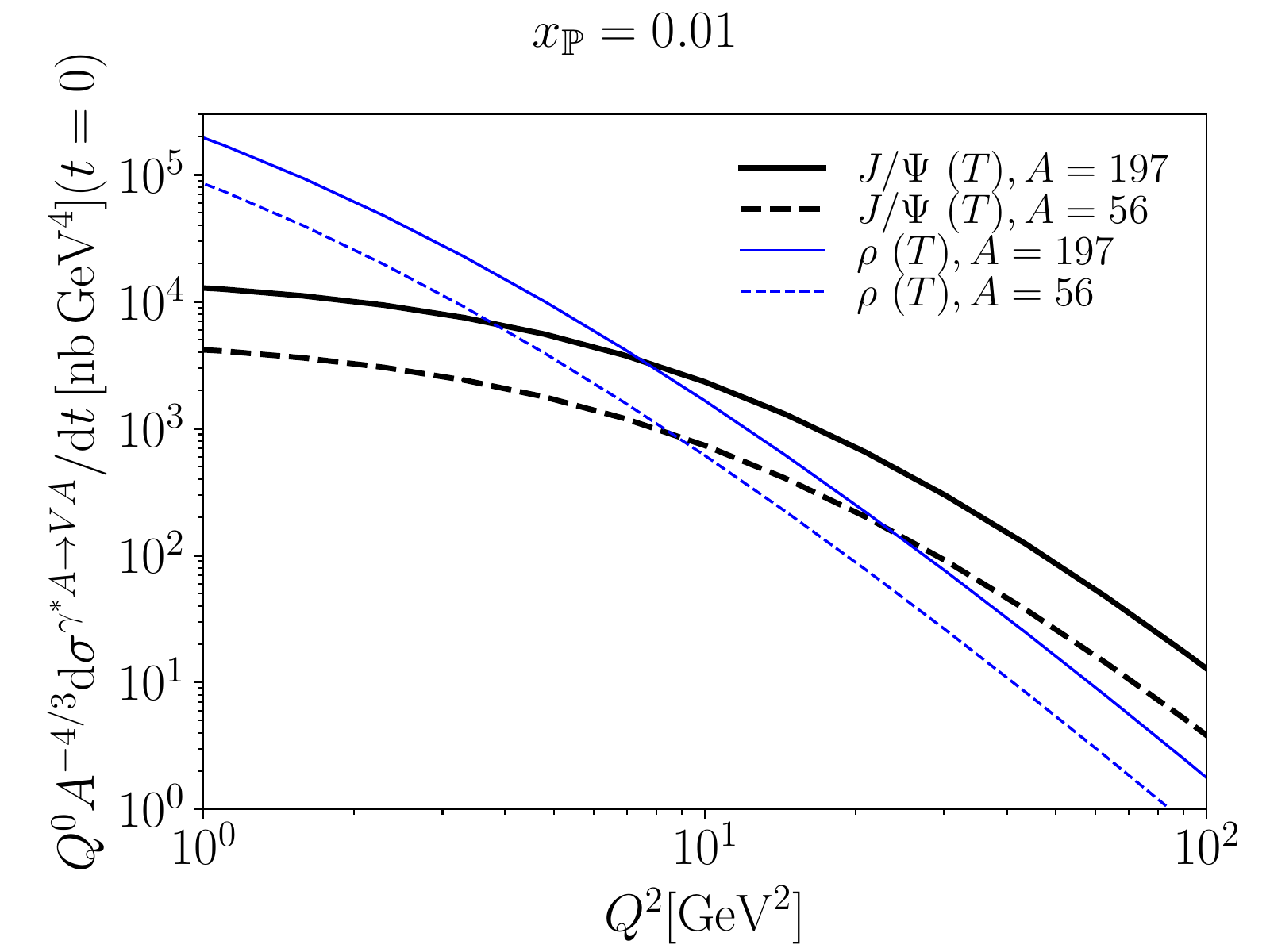} 
				\caption{The cross-section for coherent transverse vector meson production production at $t=0$. It is $Q^2$ independent at low $Q^2$. The cross-section is scaled in $A$ by  the asymptotic analytical expectation $\approx A^{4/3}$.}
		\label{fig:lowQ_scaling_Qdep_T}
\end{figure}

Deep in the saturation region, the saturation scale is much larger than $Q^2$; in these asymptotics, we can approximate $\rtsqr Q_{s,A}^2 \gg 1$. Thus the dipole-nucleus cross-section in this limit is given by $\der \sigmaa/\der^2 \bt = 2$, and the diffractive scattering amplitude in  \eq~\eqref{eq:diffamp} becomes
\begin{equation}
\A = i \int \der^2 \bt\, \der^2 \rt  \frac{\der z}{4\pi} \Psi_{\gamma^*\to q\bar q}\Psi^*_{q\bar q \to VM} \times 2 \, .
\label{eq:lowQ-amp}
\end{equation}

The $Q^2$ and $A$ dependence of  Eq.~\eqref{eq:lowQ-amp} are determined by the scale of the dipole radius and how that influences the overlap of wavefunctions. We expect that the dipole cross-section becomes independent of $r$ for $r\geq 1/Q_{s,A} \sim A^{-1/6}$. The overlap of wavefunctions in 
Eq.~\eqref{eq:lowQ-amp} for longitudinally polarized virtual photons is then  
\begin{equation}
\label{eq:overlap_smallq}
\A_L \sim \int \der r r\, \Psi_{\gamma^*\to q\bar q}\Psi^*_{q\bar q \to VM} \sim \int \der r r \,Q K_0(\varepsilon r)\, .
\end{equation}
In order to obtain some intuition as to what happens when $Q^2 \ll Q_s^2$, we will assume\footnote{In our numerical computations we take  $m_q=0.14\gev$ for $\rho$ production and $m_c=1.4 \gev$ for $J/\Psi$ production used in Ref.~\cite{Kowalski:2006hc} to obtain the wave function parametrizations.} that $Q \ll m_q$.  We emphasize that due to the small mass of the light vector mesons, the results  at small $Q^2$ are at the edge of the applicability of our weak coupling framework.

The scattering amplitude in these asymptotics has the form 
\begin{equation}
\label{eq:smallq_wavef_k}
	\A_L \sim Q \int \der r r K_0(m_q r) \approx Q \int_{1/Q_{s,A}}^{1/m_q} \der r r K_0(m_q r)\,.
\end{equation}
We will employ the identity
\begin{multline}
	\int_c^1 \der x x K_0(x) = [1 - K_1(1)] + \frac{1}{2} c^2 \ln c^2 \\
	+ \frac{1}{4}(-1 + \gamma_E - 2\ln 2)c^2 + \mathcal{O}(c^4) \, ,
\end{multline}
where $c = m_q/Q_{s,A}$ or another small number of the order of $Q^2/Q_{s,A}^2, M_V^2/Q_{s,A}^2$ in the limit of low $Q^2$. The constant term dominates at small $c$;  we therefore obtain 
\begin{multline}
	\A_L \sim Q \times \text{const} + Q \times \mathcal{O}(m_q^2/Q_{s,A}^2, M_V^2/Q_s^2, Q^2/Q_{s,A}^2)\,.
\end{multline}
 It is reasonable to anticipate that the non-perturbative scale is $r\lesssim 1/M_V$. This sets the upper limit to $1/M_V$, and thus $c=M_V/Q_s$.

Since the constant dominates in the above expression, the only $Q$ dependence we are left with is the overall $Q$ scale from the virtual photon wavefunction. Hence the exclusive vector meson cross-section goes as 
\begin{equation}
\frac{\der \sigma_L^{\gamma^*A \to VA}}{\der t} \sim Q^2
\end{equation}
 for $Q_{s,A}^2 > Q^2$. This is demonstrated in Fig.~\ref{fig:lowQ_scaling_Qdep}, where the coherent vector meson production cross-section at $t=0$ is shown as a function of $Q^2$, scaled with $Q^{-2}$. The flattening of the obtained at low $Q^2$ demonstrates the $Q^2$ scaling. However, we note that this flattening for $\rho$ takes place only at $Q^2 \lesssim 0.1 \gev^2$, and our perturbative computation is not robust in that region as discussed at the beginning of this section.   

Further, the approximation in \eq~\eqref{eq:lowQ-amp} is not justified in case of $J/\Psi$ production in realistic kinematics as its large mass and that of the charm quark limits the contribution from large dipoles to the cross-section. Thus in realistic kinematics, the $Q^2$ scaling for $J/\Psi$ production at small $Q^2$ is obtained by again linearizing the dipole cross-section. This effectively adds two powers of $r$ in \eq~\eqref{eq:smallq_wavef_k}. Noting that $\int_c^1 \der x x^3 K_0(x) = \text{const} + \mathcal{O}(c^4)$, we find exactly the same scaling at low $Q^2$ than in case of $\rho$ production, with the expectation that the low-$Q^2$ limit is reached already at $Q^2\approx M_V^2$.
 
In the case of transversally polarized photons, the vector meson overlap in \eq\eqref{eq:overlap_smallq} gives $\sim r \varepsilon K_1(\varepsilon r)$, where the extra power $r$ comes from the derivative of $\phi_T$. Approximating again $\varepsilon \approx m_q$, the diffractive scattering amplitude is proportional to 
 \begin{multline}
 \int_{1/Q_s}^{1/m_q} \der r r^2 m_q K_1(m_q r) = m_q^{-2}  \int_{c}^1 \der x x^2 K_1(x) \\
  \sim \text{const} + \mathcal{O}(c^2),
\end{multline}
where we wrote $x=m_q r$, and $c=m_q/Q_s$.
Thus in this case we do not expect to have any $Q^2$ dependence at small $Q^2$:
\begin{equation}
\frac{\der \sigma_T^{\gamma^*A \to VA}}{\der t} \sim Q^0.
\end{equation}
The scaling of the vector meson cross-sections at low $Q^2$ for transversally polarized photons is shown in Fig.~\ref{fig:lowQ_scaling_Qdep_T}.

\begin{figure}[tb]
\centering
		\includegraphics[width=0.5\textwidth]{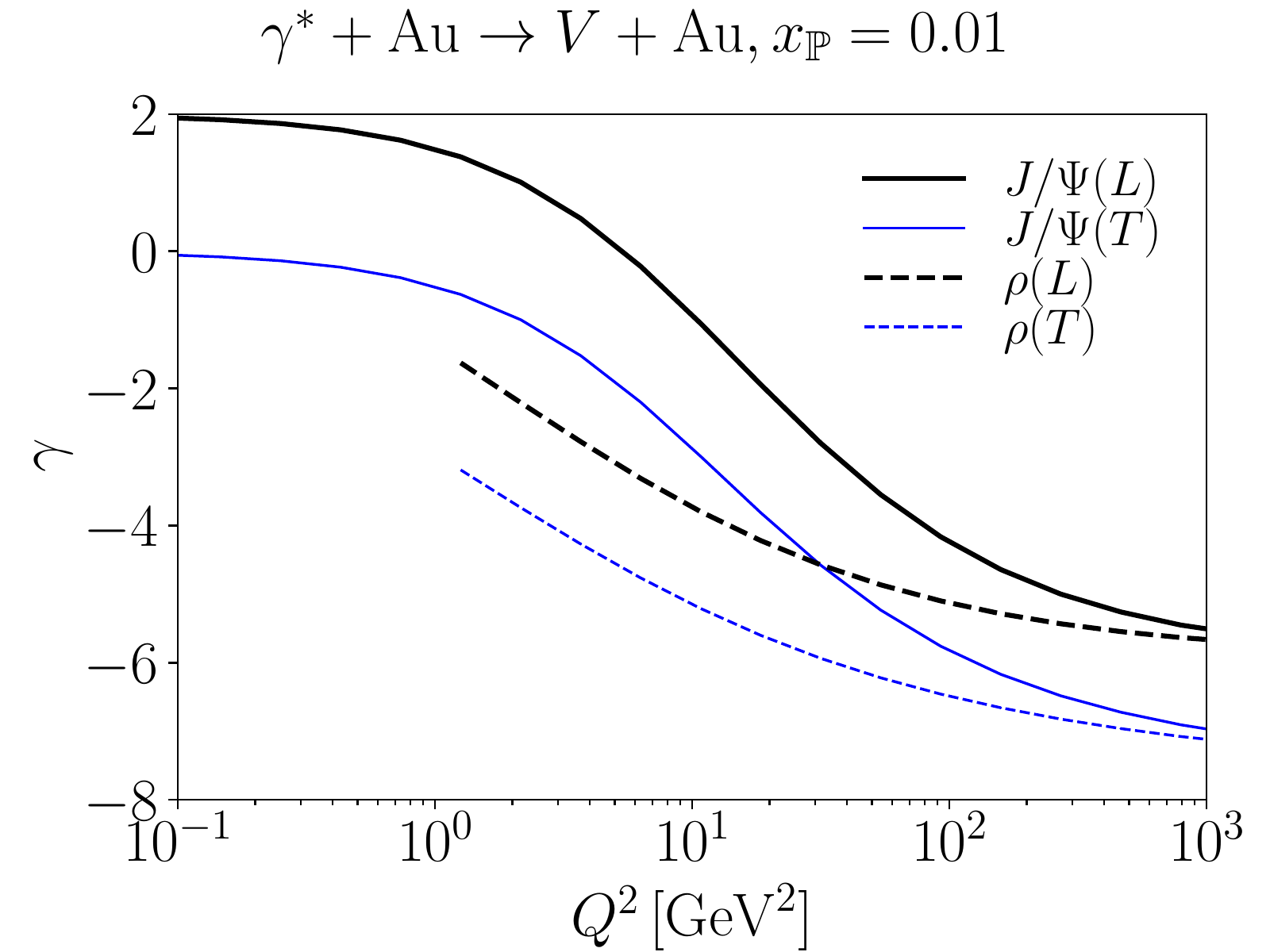} 
				\caption{Exponent of $Q$ for the exclusive vector meson production process. The result for $\rho$ production is only shown in $Q^2>1\gev^2$ where the calculation can be considered reliable.}
		\label{fig:Q2slope}
\end{figure}

We will now combine the $Q^2$ scaling results by parametrizing the exclusive vector meson cross-section as 
\begin{equation}
\sigma^{\gamma^*+A \to V+A}=c\cdot Q^\gamma,
\end{equation}
and extract the $Q$ slope parameter $\gamma$. This slope as a function of $Q^2$ is shown in Fig.~\ref{fig:Q2slope} for both $J/\Psi$ and $\rho$. The anticipated scaling changes from $Q^2$ to $Q^{-6}$ for longitudinally polarized photons, and from $Q^0$ to $Q^{-8}$ in case of transversally polarized photons. This scaling is seen clearly in the case of $J/\Psi$ mesons for $Q^2 \leq 3$ GeV$^2$. For the $\rho$, the requirement that $Q \ll m_q$ is not satisfied in the kinematical domain where we consider our framework to be reliable. The expected asymptotics for the small-$Q$ scaling exponents would be reached only at $Q^2\sim 10^{-2}\gev^2$ where our weak coupling results are clearly not trustworthy. Nevertheless, the large variation\footnote{For  instance for the $\rho$, $\gamma$ changes from $\sim -2$ to $\sim -4$ between $Q^2\sim 1$ GeV$^2$ and $Q^2=10^2$ GeV$^2$ for longitudinally polarized photons}  in $\gamma$ with $Q^2$ and the qualitatively different behavior predicted between the $\rho$ and the $J/\Psi$ cross-sections are smoking guns that will indicate whether the dynamics of gluon saturation is at play. These results will be further corroborated by the $A$ dependence which we will now turn to. 

The only dependence on $A$ in Eq.~\eqref{eq:lowQ-amp} for the low $Q^2$ region comes from the $\der^2 \bt$ integral, which gives $\int \der^2 \bt \sim A^{2/3}$. Hence for $Q_{s,A}^2 > Q^2$, one anticipates that 
\begin{equation}
\frac{\der \sigma^{\gamma^* p \to Vp}}{\der t} (t=0) \sim A^{4/3},
\end{equation}  
and the total coherent cross-section scales like $A^{2/3}$. As can be seen in Figs.~\ref{fig:lowQ_scaling_Qdep} and \ref{fig:lowQ_scaling_Qdep_T}, this asymptotic scaling regime is not reached in realistic kinematics. For example, if we look at longitudinal $\rho$ production at small $Q^2$ in the realistic kinematics of Fig.~\ref{fig:lowQ_scaling_Qdep}, the $A$ scaling turns out to be approximately $A^{1.7}$ instead of $A^{4/3}$.

\begin{figure}[tb]
\centering
		\includegraphics[width=0.5\textwidth]{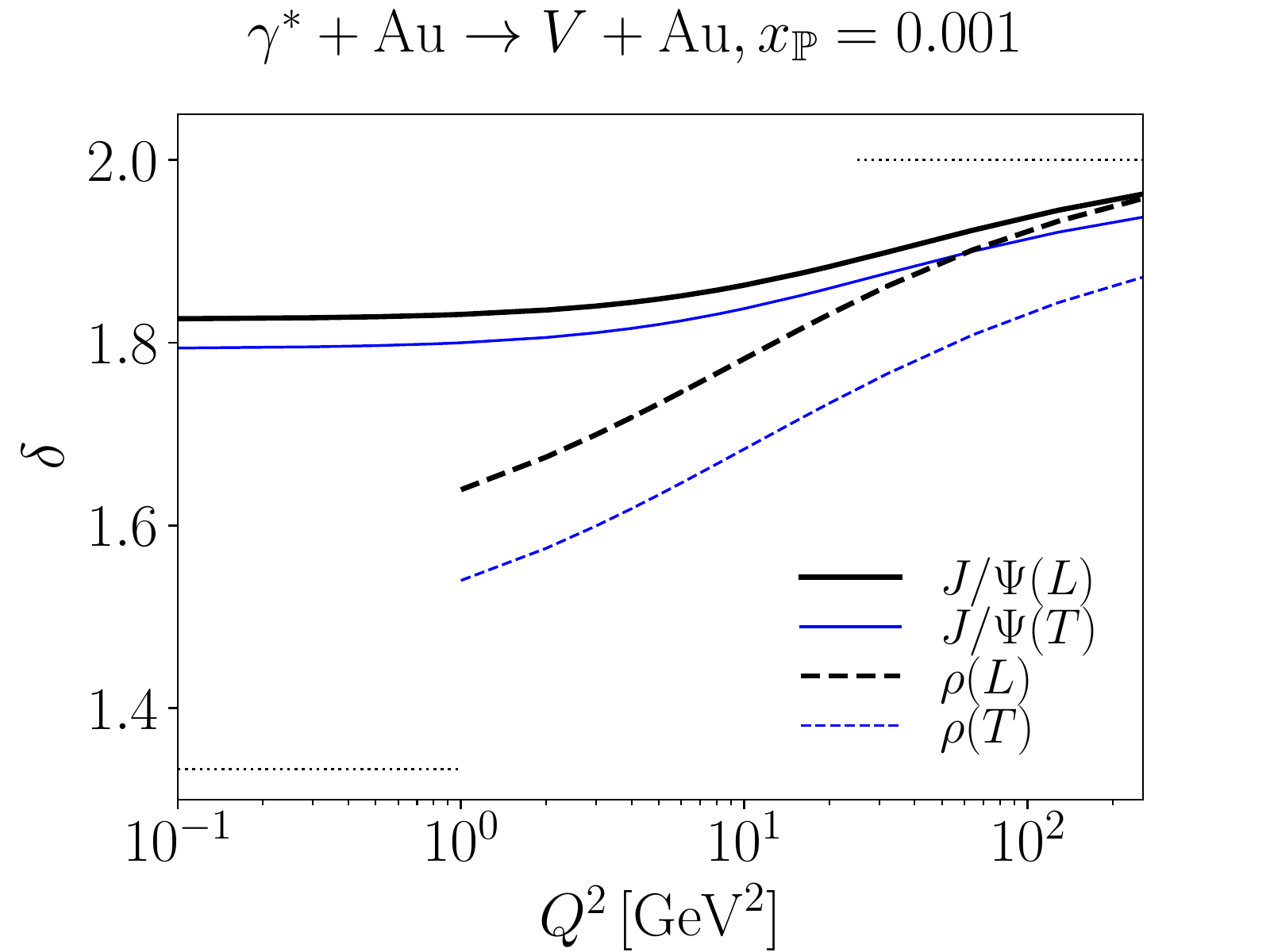} 
				\caption{Nuclear mass number $A$ scaling exponent for the coherent vector meson production at $t=0$. The cross-section is parametrized as $\der \sigma^{\gamma^* + \mathrm{Au} \to V + \mathrm{Au}}/\der t \sim A^{\delta}$.  Dotted lines show analytical results in asymptotic kinematics.}
		\label{fig:aexponent}
\end{figure}

\begin{figure}[tb]
\centering
		\includegraphics[width=0.5\textwidth]{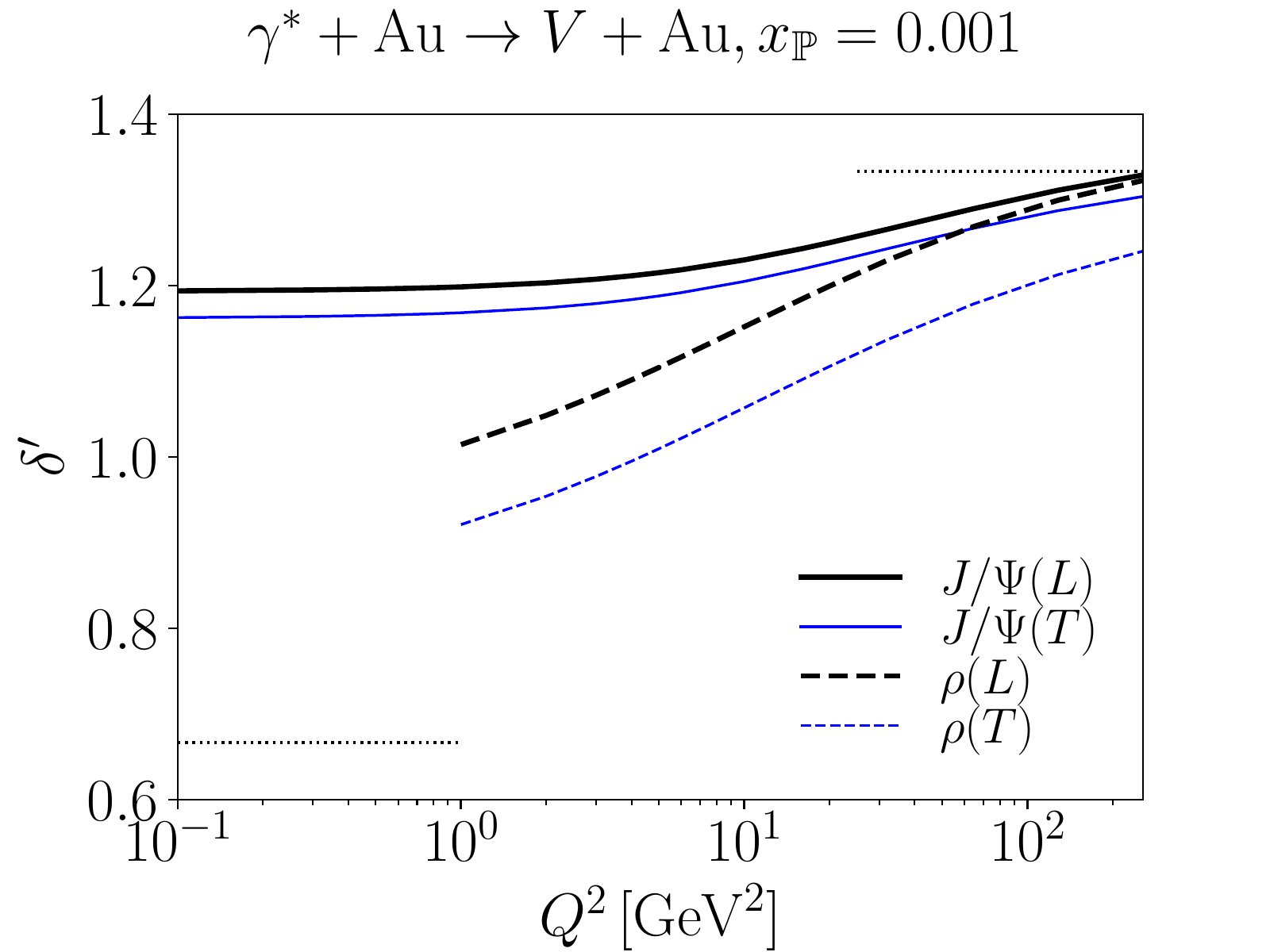} 
				\caption{Scaling exponent for total coherent vector meson production cross-section. The cross-section is parametrized as $\sigma^{\gamma^* + \mathrm{Au} \to V + \mathrm{Au}} \sim A^{\delta'}$. Dotted lines show analytical results in asymptotic kinematics. }
		\label{fig:aexponent_totxs}
\end{figure}

The numerically computed $A$ dependence in the IPsat model can be expressed as 
\begin{equation}
\frac{\der \sigma^{\gamma^* p \to Vp}}{\der t} (t=0) = c_T \times A^\delta \,.
\end{equation}  
Similarly, we can parametrize the total coherent cross-section as
\begin{equation}
\der \sigma^{\gamma^* p \to Vp} = {\tilde c}_T \times A^{\delta'}\,.
\end{equation} 
We will extract these exponents as a function of $Q^2$ for large nuclei.
These are shown for the differential cross-section in Fig.~\ref{fig:aexponent} for both $J/\Psi$ and $\rho$ at $\xpom=0.001$ (the energy dependence of the scaling exponents is weak).
Similarly, the exponents for the total coherent cross-section are shown in Fig.~\ref{fig:aexponent_totxs}.
The analytical $A^{4/3}$ scaling for $t=0$ production at low $Q^2$ is not reached in the kinematical domain accessible in the future electron-ion colliders. However the change in the $\rho$ cross-section from $\sim A^{1.5}$ to $\sim A^2$ ($\sim A^{0.9}$ to $\sim A^{4/3}$ in case of the total cross-section)  when moving from low $Q^2$ to the dilute region is observed.
In case of $J/\Psi$, at low $Q^2$ we are quite far from the analytical estimate. This is because the large mass suppresses contributions from the saturated region, and the asymptotics of $\mathrm{d}\sigmadip/\mathrm{d}^2\bt=2$  corresponding to a saturated dipole amplitude  in \eq~\eqref{eq:lowQ-amp} is not valid.

\section{Conclusions and outlook}
\begin{table*}[htp]
\begin{center}
\begin{tabular}{c|cc|cc}
 & Longitudinal, low $Q^2$ & Longitudinal, high $Q^2$ & Transverse, low $Q^2$ & Transverse, high $Q^2$ \\
 \hline
 $\der \sigma^{\gamma^* + A \to V + A}/\der t\, (t=0)$ & $ Q^2 A^{4/3}$ & $Q^{-6} A^2$ & $Q^0 A^{4/3} $& $ Q^{-8} A^2$ \\
 $\sigma^{\gamma^* + A \to V+ A}$ & $Q^2 A^{2/3} $ & $Q^{-6} A^{4/3}$ & $Q^0 A^{2/3} $ & $Q^{-8} A^{4/3}$
\end{tabular}
\end{center}
\caption{Analytical estimates for the scaling laws for the differential cross-section at $t=0$ and for the total cross-section.}
\label{table:scalings}
\end{table*}%

We demonstrated here how saturation effects significantly modify the $Q^2$ and $A$ scaling properties of the exclusive vector meson production cross-section in the cross-over from the perturbative QCD regime of large $Q^2$'s to the saturation regime of $Q^2\leq Q_{s,A}^2$. In addition to analytical estimates that are valid in asymptotic kinematics, we presented numerical results for the magnitude of these effects in the kinematics relevant for the Electron-Ion Collider.

The total cross-section for exclusive vector meson cross-section (integrated over 
$t$) at low $Q^2$ has an $A$ dependence between $A^{2/3}$ and $A$, with the numerical result closer to the latter power law dependence. This changes to $A^{4/3}$ in the pQCD regime at  $Q^2 \gg Q_{s,A}^2$. 
Similarly, as one goes from $Q^2 \gg Q_{s,A}^2$ to $Q^2 \ll Q_{s,A}^2$, from perturbative QCD to deep within the saturation regime, one observes that the  exclusive longitudinal $\rho$ meson cross-section changes its $Q^2$ dependence from $1/Q^6$ to $Q^2$ ($1/Q^8$ to $Q^0$ in case of transverse polarization). The scaling relations in asymptotic kinematics are summarised in Table~\ref{table:scalings}.
The observation of such qualitative systematics would provide strong evidence for gluon saturation.

Our calculations are performed to leading logarithmic accuracy. We note that there has been much progress in developing the saturation picture beyond  leading log accuracy~\cite{Balitsky:2008zza,Chirilli:2012jd,Iancu:2015vea,Lappi:2016fmu,Beuf:2017bpd,Ducloue:2017ftk,Hanninen:2017ddy}. In particular, exclusive vector meson production at NLO has also been computed recently~\cite{Boussarie:2016bkq}. Despite these advances, the NLO results are not  completely robust to be applied to phenomenological studies~\cite{Stasto:2013cha,Ducloue:2016shw}. It will be an important topic of future research to see if the NLO results significantly modify the results given here. Finally, we note that in addition to looking at ratios of exclusive vector meson cross-sections for different nuclei, further insight can be obtained by looking at these ratios in central $e+A$ collisions relative to those in peripheral collisions~\cite{Lappi:2014foa}.

\section*{Acknowledgements}
We thank L. McLerran and T. Sch\"{a}fer for asking the questions that inspired this note. We also thank T. Ullrich for useful discussions. 
H.M.  was supported under DOE Contract No. DE-SC0012704 and European Research Council, Grant ERC-2015-CoG-681707.
R.V's work is supported by the U.S. Department of Energy, Office of Science, Office of Nuclear Physics, under contract No. DE- SC0012704.

\bibliography{../../refs}
\bibliographystyle{JHEP-2modlong}

\end{document}